\documentclass[aps,twocolumn,amsmath,ammsymb]{revtex4-1}
\usepackage{epsfig}
\usepackage[usenames]{color}

\newcommand{\vL}{\mathcal{L}}

\newcommand{\vR}{\mathcal{R}} 

\newcommand{\vxi}{{\boldsymbol \xi} }
\newcommand{\vdelta}{{\boldsymbol \delta} }
\newcommand{\vci}{{\boldsymbol \chi} }

\newcommand{\vchi}{\mathbf{\chi}}

\newcommand{\nn}{\nonumber}

\newcommand{\dg}{\dagger}
\newcommand{\bS}{\mathbf{S}}
\newcommand{\be}{\begin{eqnarray}}
\newcommand{\ee}{\end{eqnarray}}
\newcommand{\la}{\langle}
\newcommand{\ra}{\rangle}
\newcommand{\rar}{\rightarrow}

\begin{document}

\title{Higher order spin noise spectroscopy of atomic spins in fluctuating external fields}
\author{ Fuxiang Li$^{1,2*}$,  S. A. Crooker$^3$, N. A. Sinitsyn$^{2}$}
\address{$^1$ Center for Nonlinear Studies, Los Alamos National Laboratory,  Los Alamos, NM 87545 USA}
\address{$^2$ Theoretical Division, Los Alamos National Laboratory,   Los Alamos, NM 87545 USA}
\address{$^3$ National High Magnetic Field Lab, Los Alamos National Laboratory, Los Alamos, New Mexico 87545, USA}
\email{fuxiangli@lanl.gov}

\date{\today,\now}

\begin{abstract}
We discuss the effect of external noisy magnetic fields on mesoscopic spin fluctuations that can be probed in semiconductors and atomic vapors by means of  optical spin noise spectroscopy (SNS). We show that conventional arguments of the law of large numbers do not apply to  spin correlations induced by external fields, namely, the  magnitude of the 4th order spin cumulant grows as $\sim N^2$ with the number $N$ of observed spins, i.e. it is not suppressed in comparison to the 2nd order cumulant. This allows us to design a simple experiment to measure the 4th order cumulant of spin fluctuations in an atomic system near  thermodynamic equilibrium and develop quantitative theory that explains all observations.
\end{abstract}

\date{\today}
\maketitle

\section{Introduction}

Optical spin noise spectroscopy  (SNS) \cite{zapasskii2013spin} is a minimally invasive route towards obtaining dynamical information about semiconductors \cite{oestreich2005spin,  crooker2009spin,  li2012intrinsic} and atomic gases \cite{aleksandrov1981magnetic, crooker2004spectroscopy} by measuring  time-dependent spin fluctuations.
Most of the experimental and theoretical studies have been focused so far on the  second order spin correlator  
\begin{equation}
g_2(t)=\la S_z(t) S_z(0) \ra,
\label{c2-22}
\end{equation}
where $S_z(t)$ is the time-dependent spin polarization in a mesoscopic region;  $z$ is the measurement axis, and averaging is over repeated measurements during  time intervals of duration  $T_m$, which is much larger than the characteristic spin relaxation time $\tau$. It is usually more convenient to work with the Fourier transform of this correlator, which is called the {\it  noise power spectrum}.
Let
\begin{equation}
\rho (\omega) =\frac{1}{\sqrt{T_m}} \int_0^{T_m} e^{i\omega t} S_z(t) \,dt ,
\label{aom}
\end{equation}
be the Fourier transform of a recorded spin fluctuations $S_z(t)$ during time interval $T_m$.    The {  noise power spectrum} then reads
\begin{equation}
C_2(\omega) \equiv 2\int_0^{\infty} \cos(\omega t) g_2(t) \, dt= \langle |\rho (\omega)|^2 \rangle.
\label{noiseP1}
\end{equation}

It has been argued previously  that $C_2(\omega)$ contains only restricted information about a spin system \cite{li2013higher}.
Hence, Refs.~\cite{behtold2015Finger, li2015higher, chandra2013spin, starosielec2010two, ubbelohde2012measurement, schad2014nonequilibrium, li2006many, mukamel-2D,stone2009two} proposed to explore various forms of higher order spin correlators.  The next in complexity after $C_2(\omega)$ is the third order correlator that, however, is  identically zero at thermodynamic equilibrium in nonmagnetic systems \cite{emary2007frequency, marcos2010finite,  li2013higher}.
The simplest nonzero cumulant which does not depend trivially on $C_2(\omega)$ is then the 4th order correlator or ``bi-spectrum" \cite{hinich2005normalizing}:
\begin{equation}
 C_4 (\omega_1,\omega_2) \equiv \langle |\rho (\omega_1)|^2 |\rho (\omega_2)|^2 \rangle -  \langle |\rho (\omega_1)|^2  \rangle \langle |\rho (\omega_2)|^2 \rangle.
\label{c4-1}
\end{equation}
$C_4(\omega_1,\omega_2)$ indicates how changes in the measured noise power at frequency $\omega_1$ are correlated with changes in the noise power at a different frequency $\omega_2$. Values $C_4(\omega_1,\omega_2)$ greater than zero indicate positive correlations between $|\rho(\omega_1)|^2$ and $|\rho(\omega_2)|^2$; for example, when both increase (or decrease) in synchrony.  Conversely, values of $C_4(\omega_1,\omega_2)$ less than zero reveal anti-correlations between $|\rho(\omega_1)|^2$ and $|\rho(\omega_2)|^2$; \emph{e.g.}, when an increase in $|\rho(\omega_1)|^2$ is statistically more often accompanied by a decrease in $|\rho(\omega_2)|^2$ during the same measurement time interval.


In  \cite{li2013higher}, Li {\it et al} used the stochastic path integral  technique \cite{pilgram2003stochastic} to calculate higher order correlators in Ising spin systems. They showed that the correlator $C_4(\omega_1,\omega_2)$, indeed, can reveal considerable useful information. This is also expected to be true for non-Ising spins. However, the  theory of higher order correlators  for such applications is yet to be developed. 
What is expected to be general from observations in  \cite{li2013higher}  is that spin fluctuations are practically Gaussian in  any mesoscopic volume with a large number $N$ of sufficiently weakly interacting spins, as it is articulated by the law of large numbers. By ``sufficiently weakly interacting" we mean here that dynamics of any given spin depends essentially on the state of only a few (i.e. $\ll N$) other spins at any moment of time. 
Both $C_2(\omega)$ and  $C_4 (\omega_1,\omega_2)$ then depend linearly on $N$, so that the characteristic dimensionless ratio $\eta = T_mC_4 (\omega_1,\omega_2) /[C_2(\omega_1) C_2(\omega_2)]$ is suppressed as $1/N$. This means, in particular, that to filter the 4th order cumulant from the background Gaussian fluctuations one should spend time of order $N^2$ longer than the time needed to measure the noise power $C_2(\omega)$. At the current stage, this is a very challenging task for some of the most popular applications of SNS, such as conduction electrons in semiconductors and atomic spins of warm atomic vapors, although  measurements of $C_4$ should be achievable in other potential applications of SNS such as interacting magnetic grains \cite{balk2014critical} or an  artificial  spin ice \cite{nisoli2013colloquium}.

The goal of this article is to explore the effect of an external noisy magnetic field on spin correlators $C_2(\omega)$ and $C_4(\omega_1,\omega_2)$.
 Our major finding is that even a weak noisy magnetic field induces non-Gaussian correlations with  $C_4 (\omega_1,\omega_2) \sim N^2$, i.e. higher order spin correlations
 can be experimentally observed as readily as $C_2(\omega)$.
To understand those correlations theoretically, we extend the stochastic path integral technique developed in \cite{li2013higher} to non-Ising spin systems. We find that path integrals are particularly suitable to study effects of noisy fields because averaging over the external field noise then can be done inside the path integral action prior to calculations of a specific cumulant. As higher order spin noise spectroscopy is at its very early stage of development, an additional goal of this work is to provide a simple illustrative example of a system whose higher order spin correlators can be relatively easily obtained experimentally and understood theoretically.

\section{Experiment}
We begin our discussion with a simple illustrative experiment, in which we measure the 2nd-order and 4th-order correlators of the intrinsic spin fluctuations ($C_2$ and $C_4$) from a warm atomic vapor of atomic alkali atoms ($^{41}$K) in thermal equilibrium.  The spin noise power spectrum of $^{41}$K [\emph{i.e.}, the 2nd correlator $C_2(\omega)$] has been measured previously and is well understood \cite{mihaila2006quantitative, glasenapp2014spin}. A schematic of the experiment is shown in Fig. 1(a) and is described briefly as follows: Linearly polarized light from a continuous-wave Ti:sapphire laser is detuned by $\sim 40$~GHz from the fundamental D1 optical transition of $^{41}$K ($4S_{1/2} \leftrightarrow 4P_{1/2}$; 770.1~nm) and is weakly focused through a 10~mm long glass cell containing isotopically-enriched $^{41}$K metal and 10~Torr of nitrogen buffer gas. Heating to 185 $^{\circ}$C gives a classical vapor of $^{41}$K atoms with density $\sim 7 \times 10^{13}$/cm$^3$. The large laser detuning exceeds any Doppler or pressure broadening of the D1 absorption linewidth, ensuring that this probe laser does not pump or excite the $^{41}$K atoms to leading order.

Stochastic spin fluctuations of the atoms' $4S$ valence electrons along the $\hat{z}$ direction, $\delta S_z(t)$, impart optical Faraday rotation (FR) fluctuations $\delta \theta_F(t)$ on the transmitted probe laser due to the optical selection rules in alkali atoms and because FR depends not on absorption but rather on the left- and right-circularly polarized indices of refraction of the $^{41}$K vapor (which decay slowly with detuning away from the center of the D1 resonance). $\delta \theta_F(t)$ can be detected with great sensitivity using balanced photodiodes, providing a measure of the fluctuating spin noise $\delta S_z(t)$. Therefore, the probe laser functions as a passive and nonperturbing probe of the vapor's intrinsic spin fluctuations \cite{crooker2004spectroscopy,mihaila2006quantitative,katsoprinakis2007measurement, shah2010high, zapasskii2013optical}.

A magnetic field $B_{dc} + b(t)$ is applied along the transverse ($\hat{y}$) direction, as shown. $B_{dc}$ is a static magnetic field (of order 10 Gauss), and $b(t)$ is a much weaker time-dependent field.  Since no magnetic fields are applied along the measurement direction ($\hat{z}$) and because no optical pumping occurs (due to the large laser detuning), the $^{41}$K vapor remains unpolarized and in thermal equilibrium throughout the experiment (\emph{i.e.}, $\langle S_z(t)\rangle =0$).  Figure 1(c) shows a typical spin noise power spectrum from $^{41}$K vapor in a static magnetic field ({$B_{dc} \simeq 11$~G}, $b(t) =0$). The noise power spectrum consists of four discrete peaks that are due to random spin fluctuations of the $4S$ valence electrons of the $^{41}$K atoms.  These spin fluctuations are forced to precess about $B_{dc}$, which shifts the spin noise away from zero frequency and out to MHz frequencies. These four peaks can be regarded as spontaneous spin coherences between adjacent Zeeman sublevels of the $^{41}$K ground state (see level diagram in Fig. 1(b)).  Due to hyperfine coupling  between the $I=3/2$ nuclear spin and the spin-1/2 valence electron, the $^{41}$K ground state consists of two hyperfine manifolds with total spin $F=2$ and $F=1$, separated by $\Delta_{hf} = 254$~MHz.  These manifolds are further split by the applied field $B_{dc}$ into Zeeman sublevels denoted by their spin projection $m_F$. These sublevels are unequally spaced in energy due to the hyperfine coupling; as such the six allowed Zeeman coherences ($\Delta m_F = \pm 1$ magnetic dipole transitions) appear at slightly different frequencies, as observed and as labeled (I - VI).  Transitions {II} and {III} in the upper manifold are nearly degenerate with transitions {VI} and {V} in the lower manifold, and are not separately resolved. 

\begin{figure}
\scalebox{0.2}[0.2]{\includegraphics{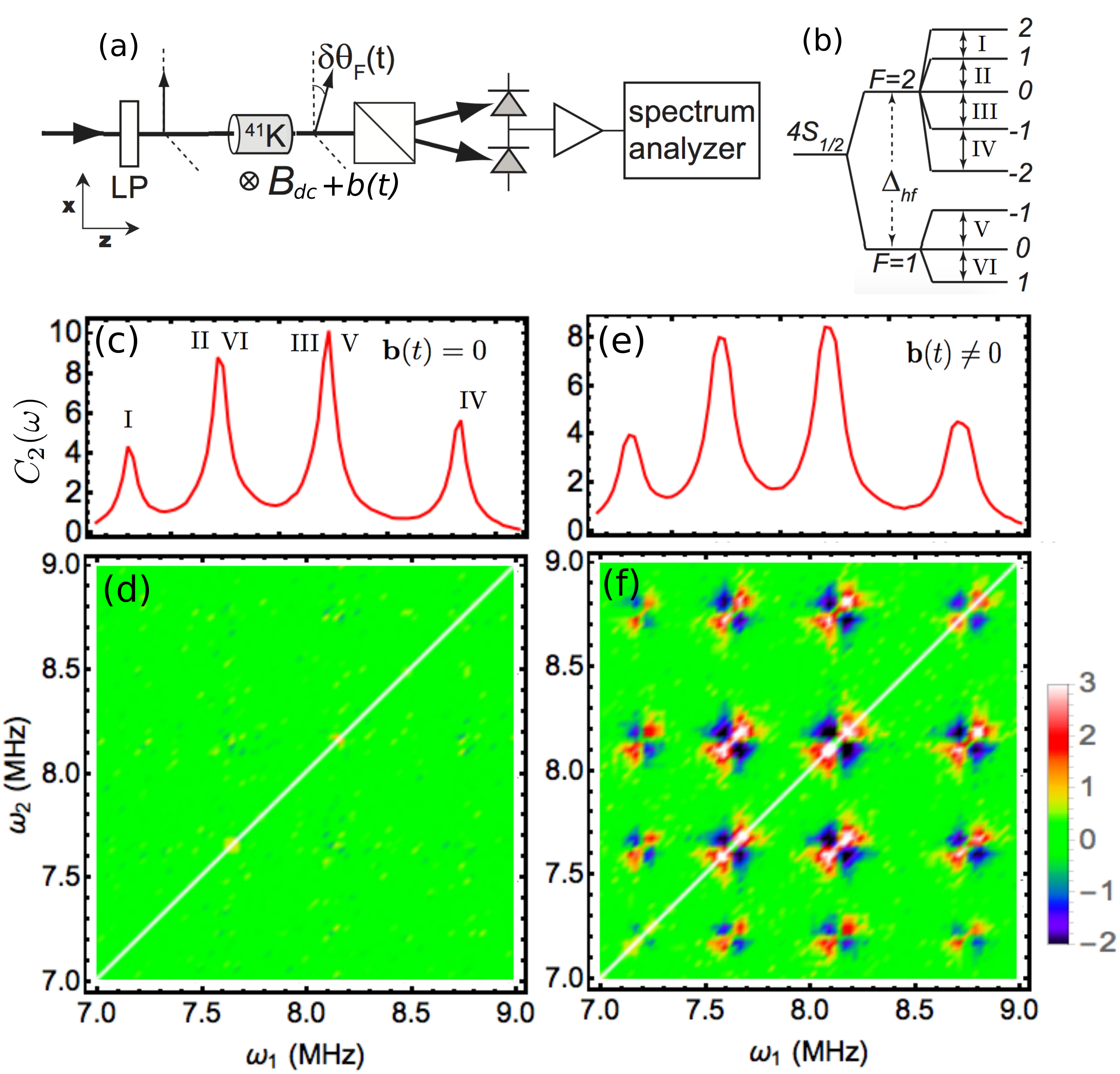}}
\hspace{-2mm}\vspace{-4mm}
\caption{ (a) Spin noise measurement setup: linearly polarized light from a tunable laser is weakly focused through a 10~mm long glass cell containing a warm $^{41}$K atomic vapor. A transverse magnetic field $B_{dc} + b(t)$ is applied. Spin fluctuations of the $4S$ valence electrons, $\delta S_z(t)$, impart Faraday rotation fluctuations, $\delta \theta_F(t)$, on the transmitted probe beam.  $\delta \theta_F(t)$ is measured and recorded directly in the time domain; from which 2nd- and 4th-order time correlators of the fluctuating spin signal are computed. (b) Diagram of the $4S$ ground state splitting of $^{41}$K atoms, induced by hyperfine coupling and a constant magnetic field.  Zeeman sublevels are labeled by their spin projection $m_F$. (c,e)  Experimentally obtained noise power spectra $C_2(\omega)$, and (d,f)  fourth order correlators $C_4(\omega_1, \omega_2)$.  In (c-d), only a static magnetic field was applied ($B_{dc}\simeq 11$~G, $b(t)=0$). In (e-f) an additional small AC field $b(t)$ was also applied, with amplitude of order 0.05~G and frequency 154~Hz. The amplitudes of $C_2(\omega)$ and $C_4(\omega_1,\omega_2)$ are in arbitrary units. In (d,f), the solid line along the diagonal $\omega_1=\omega_2$  is added for convenience as a reference line.} \label{fig:exp}
\end{figure}

When the applied magnetic field is strictly static and $b(t)=0$, the four spin noise peaks do not vary in intensity or in spectral position, and the 4th order spin noise correlator $C_4(\omega_1, \omega_2)$ does not show any structure except random background fluctuations that can be further reduced by increased signal averaging.

However, when the applied field fluctuates slightly and varies with time ($b(t) \neq 0$), $C_4(\omega_1, \omega_2)$ does exhibit a marked structure. We mimicked the effect of a fluctuating external field by applying a small additional AC magnetic field $b(t)$ along $\hat{y}$. We made the measurement interval duration $T_m$ incommensurable with the period of the AC field, so that
averaging over such time intervals  in Eq.~(\ref{noiseP1}) and (\ref{c4-1}) leads  to averaging over the phase of the AC field.

Fig.~\ref{fig:exp}(e) shows the effect of $b(t)$ on the average noise power spectrum $C_2(\omega)$. This small additional AC field shifts the four noise peaks slightly higher and lower in frequency, giving a small additional inhomogeneous broadening of the four noise peaks, but does not change or shift the noise power spectrum qualitatively.  Without prior knowledge of the intrinsic widths of the peaks, it would be difficult to tell whether the spectrum is influenced by an external noisy field or not. In contrast, the density plot of the 4th-order correlator $C_4(\omega_1, \omega_2)$ (Fig.~\ref{fig:exp}(f)) reveals a complex pattern of positive and negative correlations between different frequencies. Along the main diagonal ($\omega_1=\omega_2$), there are four bow-tie-like features, each consisting of two peaks (positive values) along the main diagonal and two valleys (negative values) along the direction transverse to the main diagonal. Moreover, Fig.~\ref{fig:exp}(f) shows that similar features also emerge away from the main diagonal, at frequency points ($\omega_i,\omega_j$), $i,j=1,\ldots,4$, where $\omega_j$ is the frequency of the maximum of the $j$-th noise power peak in Fig.~\ref{fig:exp}(c). The appearance of such off-diagonal features indicates the existence of cross-correlations between different noise resonances.


The bow-tie patterns that appear in $C_4(\omega_1, \omega_2)$ are readily understood.  Recall that $C_4(\omega_1, \omega_2)$ indicates how changes in noise power at frequency $\omega_1$ are \emph{correlated} with changes in noise power at some other frequency $\omega_2$. Figure 2 shows a simple illustrative example, where two noise power spectra are shown. The solid line shows the noise power measured in an applied transverse field $B_0$. It is peaked at frequency $\omega_L$, because the stochastic spin fluctuations in the system are forced to precess at the Larmor frequency $\omega_L = g \mu_B B_0/\hbar$.  The dotted line shows the noise power spectra measured at some later time when the field has increased slightly to $B_0 + \delta B$. The noise power has shifted to slightly higher frequencies, in accord with the slightly larger Larmor frequency.  As a result of this shift, the spin noise power for frequencies greater than $\omega_L$ has clearly \emph{increased} (note red arrows). Therefore, the noise power at a given frequency $\omega > \omega_L$ is \emph{positively correlated} with the noise power at all other frequencies greater than $\omega_L$, because the power at both frequencies increased. Similarly, the noise power for all frequencies $\omega < \omega_L$ has \emph{decreased}.  Therefore, a given frequency $\omega < \omega_L$ is also positively correlated with all other frequencies less than $\omega_L$ (because both decrease, giving a positive correlation). This is the origin of the positive values of $C_4(\omega_1, \omega_2)$ along the diagonal $\omega_1 = \omega_2$. Continuing, it is also evident that any frequency greater than $\omega_L$ is \emph{anti-correlated} with any other frequency that is less than $\omega_L$, because an increase of one is accompanied by a decrease in the other.  This is the origin of the negative $C_4(\omega_1, \omega_2)$ values that exist transverse to the diagonal. Together, these features give the characteristic bow-tie pattern of positive and negative correlations in $C_4(\omega_1, \omega_2)$, that must always be associated with a frequency-modulated signal. A related situation was discussed by Starosielec for the case of electrical noise signal in \cite{starosielec2010two}.


 \begin{figure}
\scalebox{0.5}[0.5]{\includegraphics{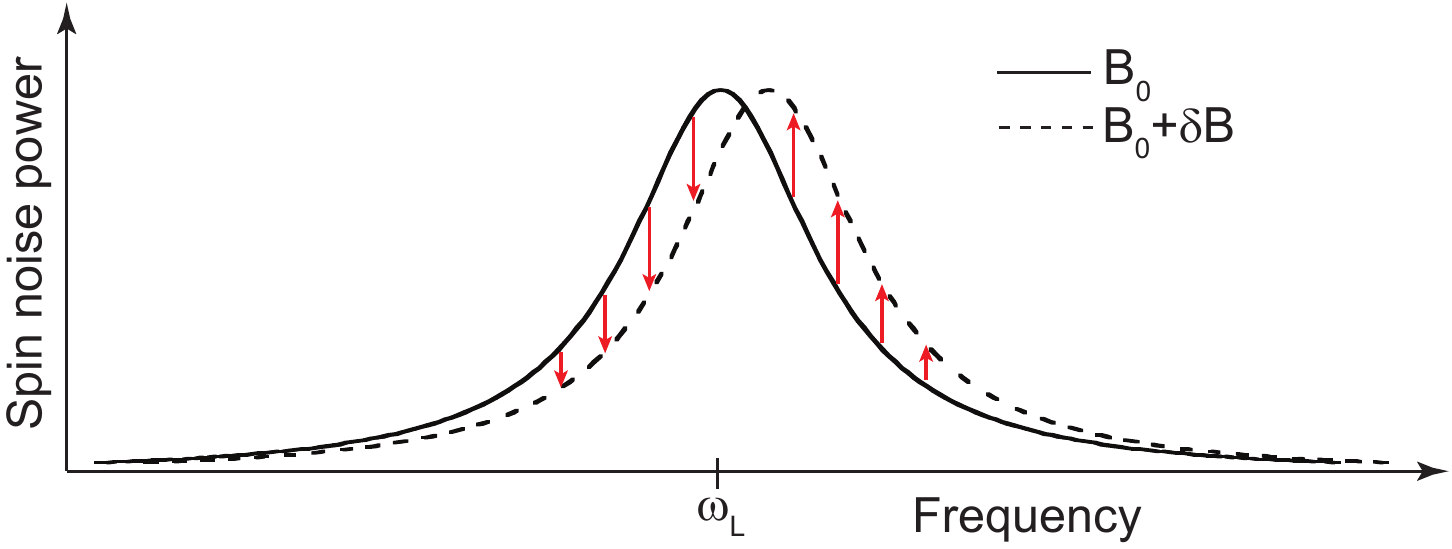}}
\hspace{-2mm}\vspace{-4mm}
\caption{A simple example to illustrate how the bow-tie patterns emerge in $C_4(\omega_1, \omega_2)$. The solid line depicts a noise power spectrum $C_2(\omega)=|\rho (\omega)|^2$ acquired in an applied field $B_0$. The spectrum is peaked at the Larmor spin precession frequency $\omega_L = g \mu_B B_0/\hbar$. The dotted line depicts a noise power spectrum acquired at some later time when the applied field is $B_0 + \delta B$.  Red arrows indicate that the measured spin noise power has increased for frequencies $\omega > \omega_L$, but decreased for frequencies $\omega < \omega_L$.  The noise power at any frequency greater than (less than) $\omega_L$ is therefore \emph{positively correlated} with all other frequencies greater than (less than) $\omega_L$, since both increased (decreased). This is the origin of positive values of $C_4(\omega_1, \omega_2)$ near the diagonal $\omega_1 = \omega_2$ (see text). Conversely, \emph{anti-correlations} exist between any frequency greater than $\omega_L$ and any frequency less than $\omega_L$, giving negative $C_4(\omega_1, \omega_2)$ along a direction transverse to the diagonal.}
\label{fig:sk}
\end{figure}

In the following, we develop the {\it quantitative} theory to interpret the patterns, such as in Fig.~\ref{fig:exp}(f), that emerge in the 4th order correlator of spins in an external noisy magnetic field.


\section{The Model and Its Stochastic Path Integral}
First, we consider the case of a weak external field, at which one can disregard the difference of g-factors of different resonances.
Dynamics of the spin polarization ${\bf S}$
of a mesoscopic volume of  a warm atomic  vapor at such conditions is well described by the  Bloch equation supplemented by a noise term $\vxi (t)$ responsible for spontaneous fluctuations of the total spin polarization:
\be
\frac{d {\bf S} }{dt}=  g  {\bf B \times S} -\gamma {\bf S} + \vxi (t),
\label{sdyn1}
\ee
where $g$ is the resonance g-factor. We absorbed the Bohr magneton in the definition of $g$, and we also set $\hbar=1$ everywhere in the text. The parameter $\gamma$ is the spin relaxation rate.
Due to the large number of atoms in the observation volume ($N\sim 10^9-10^{10}$), the intrinsic spin noise can be well approximated by a Gaussian white noise, i.e.
\be
\la \xi_{\alpha}(t) \xi_{\beta}(t') \ra = a \delta_{\alpha \beta} \delta (t-t'), \quad \alpha, \beta = x,y,z.
\label{white}
\ee
Note that, for a warm atomic vapor, the fluctuation-dissipation theorem predicts that $a \sim N \gamma$, i.e. the correlator of intrinsic spontaneous spin fluctuations is proportional to the number of spins $N$ in the observation region and the relaxation rate $\gamma$
\cite{crooker2004spectroscopy, li2013higher, roy2014cross}.

We will also assume that the constant magnetic field is applied along the y-axis:
\be
g {\bf B} = \omega_{L} \hat{\rm {\bf y}}+{\bf b}(t),
\label{mag}
\ee
where $\omega_L$ is the constant characteristic Larmor frequency, and ${\bf b}(t)$ is a stationary Gaussian noise field with an arbitrary time-correlator:
\be
\la b_{\alpha}(t) b_{\beta} (t') \ra = f_{\alpha \beta} (t-t'),
\label{mag-c}
\ee
where we assume that the field correlators $f_{\alpha \beta}(t-t')$ are arbitrary functions with positive definite power spectra in the frequency domain

The stochastic path integral is the expression for the partition function, which is the starting point for perturbative or mean field calculations of all needed cumulants of the spin noise power spectrum \cite{li2013higher}.
To derive it starting with Eq.~(\ref{sdyn1}), we write the sum over probabilities over all possible stochastic trajectories in the space  of variables ${\bf S}(t)$, $\vxi (t)$ and ${\bf b}(t)$. Each trajectory is weighted by probabilities $ P[\vxi]$ and
 $P[{\bf b}]$ of noise sources  $\vxi (t)$ and ${\bf b}(t)$, respectively. The relation between those variables given by Eq.~(\ref{sdyn1}) is included in the path integral as additional $\delta$-function weights:
\begin{widetext}
\be
Z=\int D \vxi (t) \int D {\bf b}(t) \int D{\bf S} (t) P[\vxi ]  P[ {\bf b} ]   \vdelta  \left({\bf \dot{S}}- [ g  {\bf B \times S} -\gamma {\bf S} + \vxi (t) ]  \right),
\label{z1}
\ee
\end{widetext}
where the $\vdelta$-function at a given time moment is understood as a product of three delta-functions that correspond to projections of Eq.~(\ref{sdyn1}) to different axes.
For Gaussian noise sources (\ref{white}) and (\ref{mag-c}), probabilities of their time-trajectories can be written as
\be
 &&P[\vxi ] \sim e^{-\int dt  ~ \frac { \vxi(t)^2}{2 a} }, \nn \\
  && P[{\bf b} ] \sim e^{-\iint dt ~dt'  ~  {\bf b}^T (t) \frac{ \hat{f}^{-1} (t-t')}{2} {\bf b}(t')},
\label{xi-tr}
\ee
where the elements of the matrix $\hat{f}$ are the same as  in (\ref{mag-c}).

Next, we introduce the new vector variable ${\vci}(t)$ and write the delta function as
\be
\vdelta   \left(\ldots   \right) = \int \frac{d{\vci}}{2\pi} e^{i\vci \cdot {\bf \dot{S}}  - i\vci \cdot [ g  {\bf B \times S} -\gamma {\bf S} + \vxi (t) ] }.
 \label{delta1}
\ee
After substituting (\ref{xi-tr})-(\ref{delta1}) into (\ref{z1}), we take Gaussian integrals over ${\vxi}$ and ${\bf b}$. The  partition function then reads:
\be
Z=\int D \vci (t) \int D{\bf S} (t)  e^{\vR},
\label{z2}
\ee
where $\vR =\int dt\, \vL_2(t)  + \iint dt dt'  \,  \vL_4(t,t')$, and
\be
\vL_2(t) =  i\vci \cdot {\bf \dot{S}}  - i\vci \cdot [ \omega_ {L}  {\rm {\bf \hat{y}}} \times {\bf S} -\gamma {\bf S}]  - \frac{a}{2} \vci^2,
\label{L2}
\ee
\be
\vL_4(t, t') =   -  [   \vci \times {\bf S} ]^T(t)  \frac{\hat{f}(t-t')}{2} [   {\vci \times {\bf S}} ] (t').
\label{L4}
\ee
The effect of the noisy external field is responsible for the appearance of the term $\vL_4(t, t')$ in the action $\vR$, which is of the 4th power in integration variables. Therefore, the nonperturbative regime of an arbitrary noisy field is analytically hard to treat. We will consider only weak amplitudes of time-dependent magnetic fields.

Without the noisy field term, the Lagrangian is quadratic, so we can develop a perturbative calculation approach, which is based on smallness of the correlation matrix $\hat{f}(t)$. For simplicity, we will only explore the case when the noisy field is applied along the constant field:
\be
f_{ij}(t) = \delta_{iy} \delta_{jy} f(t).
\ee
One simplification that follows then is that variables $\chi_y$ and $\rho_y$ decouple from other variables. Since we are interested in correlators of $\rho_z$, we can integrate over $\chi_y$ and $\rho_y$ and reduce the quadratic part of the action to
\be
\mathcal{L}_2 &=& i\chi_x\dot{S_x}+i\chi_z\dot{S_z} +i\gamma (\chi_x S_x+\chi_z S_z)  \nn \\
&&+ i \omega_L (\chi_x S_z- \chi_z S_x)-\frac{a}{2}(\chi_x^2+\chi_z^2).
\ee
It is straightforward, if needed, to generalize our following calculations to an arbitrary noisy field with the only price of dealing with $6\times 6$ instead of $4\times4$ matrices.

Since we are interested in cumulants (\ref{noiseP1}) and (\ref{c4-1}) in the frequency domain, it is convenient to write the path integral action
in terms of the Fourier transformed variables $\rho_{\alpha}(\omega)$ and $\chi_{\alpha}(\omega)$, where $\alpha=x,z$ and we use the convention: $\chi_{\alpha}(t)=\frac{1}{\sqrt{T_m}}\sum_{\omega}e^{i\omega t}\chi_{\alpha} (\omega)$, $\chi_{\alpha}(\omega)=\frac{1}{\sqrt{T_m}} \int dt~ e^{-i\omega t} \chi_{\alpha}(t)$.
The quadratic part of the Lagrangian in the frequency domain can then be written as
\be
\vR_2=\int dt \, \mathcal{L}_2=\sum_{\omega>0}{\bf A}^{\dg}(\omega)\hat{M}{\bf A}(\omega),
\ee
with $\quad A^{\dg}_{j} (\omega) \equiv A_{j} (-\omega)$, where
\be
{\bf A}(\omega) \equiv (\chi_x(\omega), \rho_x(\omega), \chi_z(\omega), \rho_z(\omega)),
\label{AA}
\ee
\be
\hat{M}=\left(
          \begin{array}{cccc}
            -a & -\omega+i\gamma & 0 & i \omega_L \\
            \omega+i\gamma & 0 & -i \omega_L & 0 \\
            0 & -i \omega_L & -a & -\omega+i\gamma \\
            i\omega_L & 0 & \omega+i\gamma & 0 \\
          \end{array}
        \right).
        \label{eq:M}
\ee
The quartic term in the frequency domain reads:
\begin{eqnarray}
\nonumber && \vR_4 = \iint dt \, dt'  \,  \vL_4(t,t')=-\frac{1}{T_m}\sum_{\omega} \frac{f(\omega)}{2}\sum_{\omega_1, \omega_2} \Big{[} \\
\nonumber && \chi_x(\omega_1)\rho_z(-\omega_1+\omega)\chi_x(\omega_2)\rho_z(-\omega_2-\omega) - \nn\\
\nonumber &-&2\chi_x(\omega_1)\rho_z(-\omega_1+\omega)\chi_z(\omega_2)\rho_x(-\omega_2-\omega)+\\
&+&\chi_z(\omega_1)\rho_x(-\omega_1+\omega)\chi_z(\omega_2)\rho_x(-\omega_2-\omega)\Big{]},
\label{eq:L4w}
\end{eqnarray}
where $f(\omega)=\int dt f(t) e^{i\omega t}$, and $f(t)=\frac{1}{T_m}\sum_{\omega} f(\omega) e^{-i\omega t}$.

\section{Broadening of noise power spectrum}

Let us now demonstrate how the  noisy magnetic field leads to the broadening of  the 2nd order spin correlator.
Consider the matrix $\hat{G}$ of 2nd order correlation functions: $G_{ij}= \la A_i(-\omega) A_j(\omega) \ra $, where $i, j=1,\ldots,4$ and $A_j$ are components of the vector (\ref{AA}):
\be
G_{ij}= \int \prod_{\omega'} d\vci (\omega' ) \int \prod_{\omega''} d{\bf S} (\omega'')  A_i(-\omega) A_j(\omega)  e^{\vR}, \nn\\
\label{cor21}
\ee
where we used the fact that $Z=1$  because the sum of probabilities of all possible trajectories is identically unity.
\begin{figure}[!htb]
\scalebox{0.8}[0.8]{\includegraphics{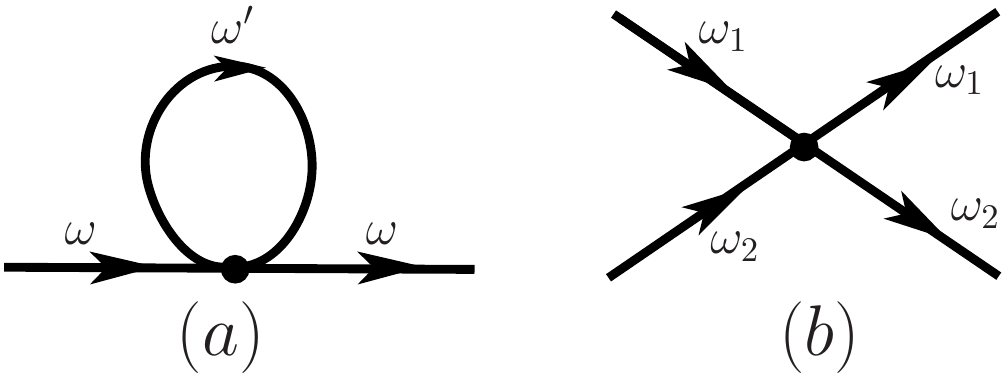}}
\hspace{-2mm}\vspace{-4mm}
\caption{ (a) The Feynman diagram contributing to the noise power spectrum. The node corresponds to multiplication by $f(\omega)$ and the loop corresponds to the propagator of the type  $\la \rho_z(-\omega') \chi_z(\omega')\ra$. Summation over the frequency of the loop is assumed.   (b) The diagram responsible for the 4th order cumulant $C_4(\omega_1, \omega_2)$.  Arrows represent the propagators  $\la A_i (-\omega) A_j (\omega) \ra$, in which external ends correspond to the measurable variables $\rho_z (\pm \omega)$ for, respectively, in-going and outgoing arrows;  the intersection point represents the coupling to the 4th order interaction term in (\ref{c4-4}), and the summation over all possible its splittings by the Wick rule. The summation over all indexes at the intersection point is assumed.}
\label{fig:feynman}
\end{figure}
Keeping only quadratic part of the action $\vR$, the correlation matrix in the absence of the noisy field can be found by performing Gaussian integration:
\be
 \hat{G}_0=-\hat{M}^{-1},
\label{inverse}
\ee
where $\hat{M}$ is written  in Eq.~(\ref{eq:M}). The general explicit expressions for elements of  $ \hat{G}_0$ are too bulky to be shown here, however, they strongly simplify
 in the experimentally relevant limit of $\omega, \omega_L \gg \gamma$:
\be
\hat{G}_0(\omega)= - \left(
          \begin{array}{cccc}
            0 &  -D_2 & 0 & - i D_2\\
            D_2^* &  D_1 & i D_2^* & i D_1 \\
            0 & i D_2 & 0 & -D_2 \\
           iD_2^*  &  -i D_1& D_2^* & D_1\\
          \end{array}
        \right),
\ee
where
\begin{equation}
D_1\equiv \frac{a/2}{(\omega- \omega_L)^2+ \gamma^2}, \quad
D_2 \equiv \frac{1}{2} \frac{1}{\omega-\omega_L +i \gamma}.
\label{propagators}
\end{equation}
For example, the noise power spectrum in the absence of a time-dependent magnetic field is given by
\be
\la |\rho (\omega)|^2 \ra_0\equiv\hat{G}_{0,22}  = \frac{a/2}{(\omega-\omega_L)^2+\gamma^2} .
\label{eq:c2}
\ee


In order to include the effect of the quartic term (\ref{eq:L4w}) we solve the Dyson equation  $\hat{G}=\hat{G}_0+\hat{G}_0 \hat{\Sigma} \hat{G}$, or equivalently
\be
\hat{G}^{-1}=\hat{G}^{-1}_0+ \hat{\Sigma}.
\label{dyson}
\ee
Comparing Eq.~(\ref{dyson}) with Eq.~(\ref{eq:M}), we  conclude that the self energy $\hat{\Sigma}(\omega)$ will renormalize both the relaxation rate $\gamma$ and the fluctuation amplitude $a$. The real part of  $\Sigma_{11}$ will renormalize $a$, while the imaginary part of the matrix element $\Sigma_{12}$ will renormalize $\gamma$.     Up to the leading order in $f(\omega)$,  the self-energy is described by the loop part of the Feynman diagram shown in Fig.~\ref{fig:feynman}(a).
From the quartic part of the action (\ref{eq:L4w}), only the second term contributes to the self-energy ${\Sigma}_{12}(\omega)$, which then corresponds to the propagator $\la \rho_i(-\omega) \chi_i(\omega)\ra$. Explicitly, 
\be
\Sigma_{12}(\omega) = \frac{1}{T_m} \sum_{\omega'} f(\omega'-\omega ) \la \rho_z(-\omega') \chi_z(\omega')\ra.
\label{eq:sigma}
\ee

For example, in the case of a Gaussian white noise correlator of the noisy field, $f(t)=f_0\delta(t)$ and $f(\omega) = f_0$, Eq.~(\ref{eq:sigma}) reduces to
 $\hat{\Sigma}_{12}(\omega) = f_0 \frac{1}{T_m} \sum_{\omega'} \la \rho_z(-\omega') \chi_z(\omega')\ra  = i f_0 /2 $. Therefore,
the  effect of the noisy field in this case is  renormalization of the  relaxation rate as
 \be
 \gamma \rar \gamma' = \gamma +\frac {f_0}{ 2 }.
 \ee
 Similar calculations lead to renormalization $a \rightarrow a\gamma'/\gamma$. The form of the noise power spectrum remains Lorentzian, as in the unperturbed case described by Eq.~(\ref{eq:c2}), which agrees qualitatively with our experimental results. Note that the total area of the noise power spectrum remains unchanged, which means that the noisy magnetic field changes dynamics of fluctuations but not  the equal time spin correlator.



\section{4th order cumulants}
At zeroth order in $f(t-t')$ the 4th order correlator is identically zero. Hence, one does not have to perform the summation of the infinite series of diagrams in order to calculate $C_4(\omega_1, \omega_2)$. At the leading
linear order in $f(t-t')$ we find
\be
C_4(\omega_1, \omega_2)=\la |\rho_z(\omega_1)|^2 |\rho_z(\omega_2)|^2 \vR_{4}\ra_0,
\label{c4-4}
\ee
where $\la ... \ra_0$ again means that the averaging is taken over the quadratic action in the path integral.

By applying the Wick's rule, one can find that all relevant Feynman diagrams have the topology of an intersection of four propagators, as shown in Fig.~\ref{fig:feynman}(b).
For example, three separate terms in Eq.~(\ref{eq:L4w}) produce the following three contributions to (\ref{c4-4}):
\begin{widetext}
\be
C_4^{(1)}&&=C_4^{(3)} =\frac{1}{T_m}\frac{a^2}{16}\frac{f(0)4\Omega_1\Omega_2+f(\Omega_1-\Omega_2)(\Omega_1+\Omega_2)^2+f(\Omega_1+\Omega_2) [(\Omega_1+\Omega_2)^2+4\gamma^2]}{(\Omega_1^2+\gamma^2)^2(\Omega_2^2+\gamma^2)^2}, \\
C_4^{(2)}&&=\frac{1}{T_m}\frac{a^2}{8}\frac{f(0)4\Omega_1\Omega_2+f(\Omega_1-\Omega_2)(\Omega_1+\Omega_2)^2-f(\Omega_1+\Omega_2) [(\Omega_1+\Omega_2)^2+4\gamma^2]}{(\Omega_1^2+\gamma^2)^2(\Omega_2^2+\gamma^2)^2},
\ee
\end{widetext}
where
\be
\Omega_{1, 2} \equiv \omega_{1,2}-\omega_L.
\ee

 Adding all those contributions, we obtain the explicit expression for the bi-spectrum in the limit of a weak noisy field:
\be
C_4=\frac{a^2}{2T_m} \frac{4 f(0) \Omega_1 \Omega_2 + f(\Omega_1-\Omega_2)(\Omega_1+\Omega_2)^2}{(\Omega_1^2+\gamma^2)^2(\Omega_2^2+\gamma^2)^2} .
\label{eq:C4f}
\ee
Here we recall that $a\sim N\gamma$, so that  $C_4\sim N^2$, which confirms that $C_4$ is not suppressed in comparison with $C_2^2$.

\begin{figure}[tb]
\scalebox{0.35}[0.35]{\includegraphics{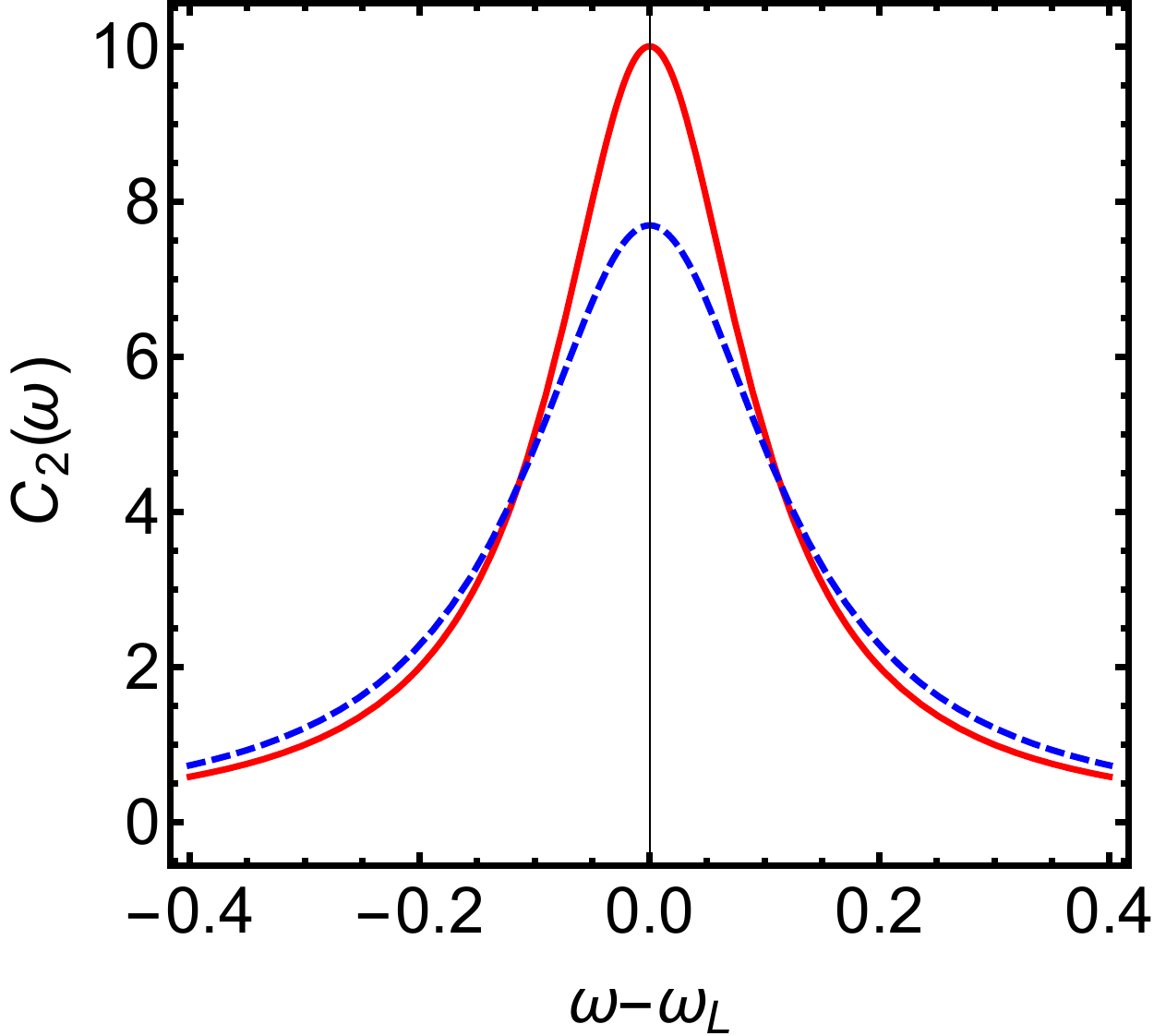}}
\hspace{-2mm}\vspace{-4mm}
\caption{The second order correlator $C_2(\omega)=\la |\rho_z(\omega)|^2 \ra$ in the absence of a noisy magnetic field (red solid line), and in presence of a noisy magnetic field (blue dashed line). The external field noise  correlator  is $f(t)=J^2e^{-\gamma_t t}$. Parameters of the model are: $\gamma=0.1$, $J=0.01$ and $\gamma_t = 0.3 $. } \label{fig:c2}
\end{figure}

\begin{figure}[t]
\scalebox{0.3}[0.3]{\includegraphics{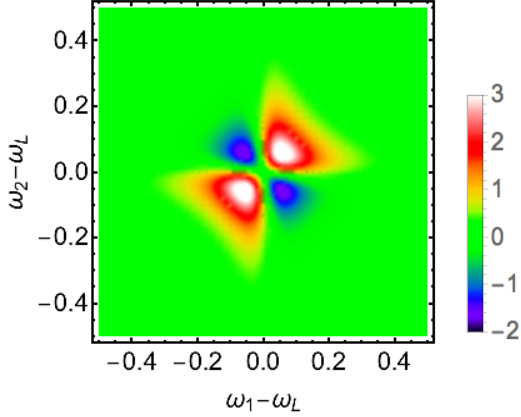}}
\hspace{-2mm}\vspace{-4mm}
\caption{ The density plot of the 4th order correlator $C_4(\omega_1, \omega_2)$, according to Eq.~(\ref{eq:C4f}).  External field noise has correlator  $f(t)=J^2e^{-\gamma_t t}$. Parameters of the model are: $\gamma=0.1$, $J=0.01$ and $\gamma_t = 0.3 $.} \label{fig:c4}
\end{figure}

Consider, for example,  a noisy field correlator $f(t)=J^2e^{-\gamma_t t}$, whose Fourier transform is $f(\omega)=J^2\frac{2\gamma_t}{\omega^2+\gamma_t^2}$. In Fig.~(\ref{fig:c2}) and Fig.~(\ref{fig:c4}), we compare the effect of such a noisy field on correlators $C_2(\omega)$ and $C_4(\omega_1, \omega_2)$. As in our experiment, $C_2(\omega)$ merely acquires additional broadening, while the 4th order correlator becomes nonzero and shows the experimentally observed pattern that consists of two peaks along the main diagonal and two valleys in transverse to the main diagonal direction.

\begin{figure}[!htb]
\scalebox{0.3}[0.3]{\includegraphics{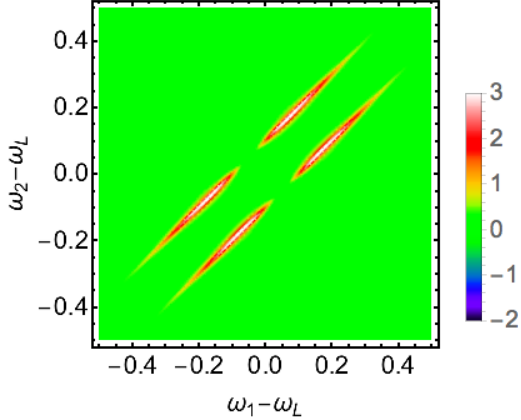}}
\hspace{-2mm}\vspace{-4mm}
\caption{Density plot of $C_4$ for the case when the noisy magnetic field power spectrum is sharply peaked at finite frequency $\omega_s = 0.1$.
Other parameters are $\omega_L=1$, $\gamma=0.2$ and $g=0.1$.  } \label{fig:c4_sinu}
\end{figure}

Another example that helps to develop the intuition about the 4th correlator behavior corresponds to the limiting case of a  time-dependent field whose correlator in the frequency domain is sharply peaked near a  frequency $\omega_s$:  $f(t) = 2 g e^{-\delta t} \cos \omega_s t$, where $\delta \ll \omega_s$.  In the frequency domain, such a correlator can be approximated by a sum of two delta-functions: $f(\omega) =g   \delta(\omega- \omega_s) + g\delta(\omega+\omega_s) $.
Figure~\ref{fig:c4_sinu} shows that the density plot of $C_4$ then  consists of four narrow regions along the lines $\Omega_1 - \Omega_2 = \pm \omega_s$. The bow-tie pattern disappears because it is specific either for quasi-static fluctuations or, at least, fluctuations with a substantial value of $f(0)$. In our experiment, the latter was finite because the AC-field oscillation period was smaller than the measurement time interval $T_m$ for statistical averages. The latter introduced a cut-off that smeared the spectrum of the external field. Note also that observation of splitting of narrow lines in Fig.~\ref{fig:c4_sinu} requires external field frequency $\omega_s$ to be comparable to the spin relaxation rate $\gamma$. In contrast, our experiment corresponded to the limit $\omega_s \ll \gamma$, so the behavior shown in Fig.~\ref{fig:c4_sinu} could not be observed at such parameters even by changing $T_m$.  

We also note that having the density plot for $C_4$ we can reconstruct the correlator of the noisy field in the frequency domain, i.e. $f(\omega)$. To show this, let us set
$\Omega_2=0$ in (\ref{eq:C4f}):
\be
C_4(\Omega, 0) = \frac{a^2}{2T_m \gamma^4} \frac{f(\Omega) \Omega^2}{(\Omega^2 +\gamma^2)^2}.
\label{recon}
\ee
Parameters $\omega_L$ and $\gamma$ can be obtained with a high precision from the knowledge of the noise power spectrum $C_2(\omega)$. Hence, one can revert Eq.~(\ref{recon}) and obtain $f(\Omega)$.




\section{Cross-Correlations and off-diagonal peaks}
Finally, we will look at the case with more than one resonance.
For simplicity, we consider a situation with only two resonances having different  g-factors, fluctuation amplitudes $a$, relaxation rates $\gamma$, and different coupling strengths to the measurement beam.

Let, in a constant external magnetic field, the positions of the noise power peak maxima be centered at frequencies $\omega_{L}$ and $\omega_L' $ and the ratio between the two peak frequencies, and consequently the ratio of g-factors, be
$$
r \equiv \omega_L'/\omega_L.
$$
 We denote variables that correspond to the first resonance by ${ \vchi}$, ${\bf S}$, $a$,  $\gamma$, and those of the second resonance by $\vci'$, $\bS'$ , $a'$ and $\gamma'$.

 The quadratic part of the  Lagrangian  consists of two copies of ${\cal L}_2$ from Eq.~(\ref{L2}), i.e., ${\cal L}^t_2 = {\cal L}_2+ {\cal L}_2'$, where  ${\cal L}_2'$ is the same as ${\cal L}_2$ with $\vci$,  $\bS$, $\omega_L$, $a$ and $\gamma$ replaced by $\vci'$,  $\bS'$, $\omega_L'$, $a'$ and $\gamma'$.
After the averaging over the magnetic field noise, the quartic part of the Lagrangian density consists of three contributions:
\be
{\cal L}_4^t = {\cal L}_4+ {\cal L}_4'+ {\cal L}_4^{c} .
\ee
The first term  ${\cal L}_4$ is the same as Eq.~\ref{L4}, while the second term ${\cal L}_4'$ is the same as ${\cal L}_4$ with $\vci$ and  $\bS$ replaced by $\vci'$ and $\bS'$, and with $f(t)$ replaced by $r^2 f(t)$. The third term is a cross-term ${\cal L}_4^c$:
\be
{\cal L}_4^{c} = -  r [   \vci \times {\bf S} ]^T_t  \hat{f}(t-t') [   {\vci' \times {\bf S}'} ]_{t'}.
\label{L4c }
\ee

\begin{figure}
\scalebox{0.28}[0.28]{\includegraphics{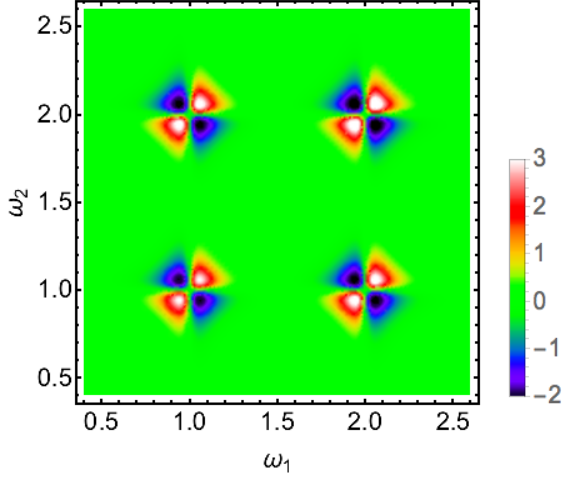}}
\hspace{-2mm}\vspace{-4mm}
\caption{Density plot of 4th order cumulant $C_4^t(\omega_1, \omega_2)$ (arb. units) for the case of two resonances with different g-factors. The Larmor frequencies of resonances are $\omega_L=1$ and $\omega_L' =2$,  the relaxation rates are the same, $\gamma=0.1$. The effects of the relative coupling $q$ to the measurement beam  and the ratio of g-factors $r$ can be combined together. Here we choose  $q\, r=1.1$.  The external field noise correlator is $f(t)=J^2e^{-\gamma_t t}$, with  $J=0.01$ and $\gamma_t = 0.1 $.} \label{fig:cross}
\end{figure}

In the case of two resonances, the measured bi-spectrum would be
\be
C_4^t(\omega_1, \omega_2) \equiv \la |\rho^t(\omega_1)|^2 |\rho^t(\omega_2)|^2 \ra - \la |\rho^t(\omega_1)|^2 \ra \la|\rho^t(\omega_2)|^2 \ra, \nn \\
\label{c4-5}
\ee
where $$\rho^t(\omega) = \rho(\omega) + q \rho' (\omega),$$  with $q$ being the relative coupling strength of the second resonance to the measurement beam.
Analogous calculations to the case with a single resonance lead to the expression for the bi-spectrum  that contains three contributions:
\be
C_4^t= C_4 + q^2 C_4' + q C_4^c,
\ee
where $C_4'$ has the same form as $C_4$ in (\ref{eq:C4f}) but with $\omega_L$ and $f(\omega)$ replaced with $\omega_L'$ and $r^2 f(\omega)$, respectively.  The third term is qualitatively different. It describes  cross-correlations induced by the coupling of both resonances to the same noisy signal. Explicitly, we find
\begin{widetext}
\be
C_4^{c} =&& 2 \frac{a a'}{T_m} r f(0) \Big[ \frac{\Omega'_1 \Omega_2}{ [(\Omega'_1)^2 +(\gamma')^2 )^2 (\Omega_2^2 +\gamma^2]^2}  + \frac{\Omega_1 \Omega'_2}{[\Omega_1^2 +\gamma^2 ]^2 [(\Omega'_2)^2 +(\gamma')^2]^2 }\Big] \nn \\
&&+ \frac{aa'}{T_m} r f(\Omega_1-\Omega_2) \frac{(\Omega_1 +\Omega_2)(\Omega'_1 +\Omega'_2)}{(\Omega_1^2+\gamma^2)(\Omega_2^2+\gamma^2)[(\Omega'_1)^2+ (\gamma')^2] [(\Omega'_2)^2+(\gamma')^2]},
\ee
\end{widetext}
where, $\Omega_{1,2} \equiv \omega_{1,2}-\omega_L$ and $\Omega_{1,2}' \equiv \omega_{1,2} -\omega_L'$.
The  plot of $C_4^t (\omega_1, \omega_2)$ is shown in Fig.~\ref{fig:cross} for the case of an exponential correlator of the external magnetic field noise.

Fig.~\ref{fig:cross} shows that, in addition to the previously discussed features along the main diagonal, cross-correlations lead to the appearance of additional off-diagonal peaks. This phenomenon is analogous to the appearance of off-diagonal peaks in two-dimensional nuclear and electronic spin resonance techniques \cite{mukamel-2D}.


\section{Discussion}

We explored the behavior of spin fluctuations under the action of an external fluctuating magnetic field. While the effect of  a weak noisy field on the 2nd order spin correlator is small, the effect on the 4th order correlator is dramatic.
Although the field is weak, it acts simultaneously on all observable spins so that dynamics of all spins become correlated. As a result, the bi-spectrum is not suppressed. The correlator $C_4(\omega_1,\omega_2)$ generally shows a complex pattern in the density plot. Such a pattern can be used to extract the time-correlator of the noisy magnetic field. This sensitivity of $C_4(\omega_1,\omega_2)$ demonstrates its potential for applications in which electronic spin noise is used to probe fluctuations of magnetic fields that originate from another system in contact with probed electrons. For example, recently,  the spin noise of conduction electrons was used to monitor dynamics of a nuclear spin polarization in semiconductors \cite{glazov-nuclear}, and spectroscopy of NV centers was used to probe local magnetic field fluctuations from surface electron spins in  diamond \cite{nV-noise}. Our results suggest that measurements of the bi-spectrum can be used as an alternative approach to explore similar noisy magnetic field sources.

The simplicity with which we induced considerable higher order correlations also means that even an extremely weak extrinsic noise can produce a detrimental effect on efforts to measure  the 4th order correlator of the intrinsic  spin fluctuations that are induced by microscopic spin interactions rather than external fields. The intrinsic spin noise is generally in the domain of application of the law of large numbers, which makes intrinsic fluctuations almost Gaussian. Our results show that without a special isolation from external field fluctuations, as well as other sources of nonlinearities in the detector, the 4th order correlators of signals from a large number of spins can be strongly distorted by extrinsic effects. We note, however, that similar effects of the noisy fields do not emerge on the level of the 3rd order spin correlator \cite{li2013higher}, which becomes an interesting alternative for the higher order SNS studies.  The relative role of extrinsic effects on the bispectrum can also be minimized in the case of strong intrinsic spin correlations, for example, in magnetization fluctuations near a phase transition.

\bibliographystyle{apsrev4-1}

\end{document}